\documentclass[aps,prx,twocolumn,floatfix,superscriptaddress,showpacs,amsmath,amssymb]{revtex4-2}
\usepackage{graphicx}
\usepackage{epsfig}
\usepackage{hyperref}
\usepackage{color,colordvi}
\usepackage{float}

\usepackage[sectionbib]{bibunits}
\defaultbibliographystyle{apsrev4-2} 
\defaultbibliography{references}

\newcommand{\aver}[1]{\langle {#1} \rangle}
\newcommand{\es}[1]{\begin{split}#1\end{split}}
\newcommand{\bea}{\begin{align}}
\newcommand{\eea}{\end{align}}

\newcommand{\lp}{\left(}
\newcommand{\rp}{\right)}
\newcommand{\lsq}{\left[}
\newcommand{\rsq}{\right]}
\newcommand{\lbr}{\left\lbrace}
\newcommand{\rbr}{\right\rbrace}
\newcommand{\da}{\dagger}
\newcommand{\bma}{\begin{pmatrix}}
\newcommand{\ema}{\end{pmatrix}}
\newcommand{\bra}[1]{\langle #1 |}
\newcommand{\ket}[1]{| #1 \rangle}
\newcommand{\mo}{{-1}}
\newcommand{\rw}{\rightarrow}
\newcommand{\oh}{\frac{1}{2}}
\newcommand{\w}{\omega}

\newcommand{\tr}{\text{tr}}
\newcommand{\abs}[1]{ \left\lvert #1	\right\rvert}
\usepackage{bbold}
\newcommand{\id}{\mathbb{1}}
\newcommand{\hc}{\text{hc}}

\newcommand{\ii}{{\rm i}}

\begin{document}
\begin{bibunit}
    
\title{Fate of the Mollow triplet in strongly-coupled atomic arrays}

\author{Orazio Scarlatella}
\author{Nigel R. Cooper}
\affiliation{T.C.M. Group, Cavendish Laboratory, University of Cambridge, J.J. Thomson Avenue, Cambridge CB3 0HE, UK\looseness=-1}


\begin{abstract}
Subwavelength arrays of quantum two-level emitters have emerged as an interesting platform displaying prominent collective effects that can be harnessed for applications. 
Here we study such arrays under strong coherent driving, realizing an open quantum many-body problem in a strongly non-linear regime. 
{For this we introduce a novel approach to this problem in terms of a Dynamical Mean Field Theory (DMFT), paving the way for further studies.}
We show that the spectrum of scattered light, characterized by the famous Mollow triplet for a single atom, develops a 
characteristic lineshape with flat sidebands determined by dipolar interactions {and relevant for experiments}. 
Remarkably, this is {to some extent }independent of the specific geometry, but is sensitive to the ordered arrangement of the atoms. 
This lineshape therefore characterizes atomic arrays and distinguishes them from disordered ensembles and non-interacting emitters. 
\end{abstract}

\maketitle



The collective behaviour of ensembles of quantum emitters, resulting from photons scattering between them, has long attracted attention \cite{agarwalAgarwal1970, dickeDicke1954, lehmbergLehmberg1970}. 
Considering two-level atoms, when the average inter-atomic distance becomes comparable to the wavelength of the two-level transition, the resulting dipolar interactions become strong and can lead to collective effects, such as superradiance and subradiance \cite{dickeDicke1954, grossHaroche1982a} and Lamb shifts \cite{friedbergManassah1973,keaveneyAdams2012}.

Cold atoms systems are natural places where strong dipolar interactions can be studied and exploited. 
There, the emitters can be arranged in ordered configurations, allowing one to enhance and control collective effects \cite{ruostekoskiRuostekoski2023}.
Such arrays of quantum emitters 
have recently been investigated for several applications  
in quantum information and optics \cite{pattiYelin2021,solntsevKivshar2021,bekensteinLukin2020}, 
including realizing single atomic layer mirrors 
\cite{bettlesAdams2016,shahmoonYelin2017,ruiBloch2020} with coherent control \cite{srakaewZeiher2023,bekensteinLukin2020}, as well as for photon storage \cite{cechOlmos2023,asenjo-garciaChang2017,facchinettiRuostekoski2016}. 
Several effects have been studied including the shift of collective resonances \cite{glicensteinBrowaeys2020,jenkinsBrowaeys2016},
electromagnetically induced transparency \cite{fengScully2017,bettlesAdams2015,srakaewZeiher2023} and superradiance and subradiance \cite{ostermannYelin2023, rubies-bigordaYelin2022a,massonAsenjo-Garcia2022,rubies-bigordaYelin2021,asenjo-garciaChang2017}.

Nevertheless, most works so far focused on weak drive or single excitation regimes, in which the emitters are described by non-interacting theories \cite{williamsonRuostekoski2020,porrasCirac2008,asenjo-garciaChang2017}. Less is known in non-linear regimes, where the system constitutes a strongly-interacting far-from-equilibrium quantum many-body problem.  

A typical non-linear effect in quantum optics characterizes the spectrum of light scattered by a single two-level atom that is strongly driven on resonance. This has a three-peak structure known as the Mollow triplet \cite{mollowMollow1969}, due to a dressing of the atomic levels with drive photons \cite{cohen-tannoudjiReynaud1977}. It was observed in several platforms including atomic beans \cite{schramaHeideman1992}, ions \cite{stalgiesToschek1996}, molecules \cite{wriggeSandoghdar2008}, quantum dots \cite{flaggShih2009} and cold atoms \cite{ortiz-gutierrezFouche2019}. 
A number of works investigated how this is affected by dipolar interactions in the case of few atoms \cite{agarwalVetri1977,senitzkySenitzky1978, carmichaelCarmichael1979,freedhoffFreedhoff1979, kilinKilin1980,vivas-vianaSanchezMunoz2021} and more recently for disordered clouds \cite{ottKaiser2013,pucciBachelard2017,jonesOlmos2017}. However its fate for large arrays of strongly-coupled atoms is still unclear, and was only investigated in small systems \cite{jonesOlmos2017,moniriDarsheshdar2023}. 

Here we show that for large arrays of strongly-coupled atoms the Mollow triplet acquires a characteristic lineshape with flat sidebands determined by dipolar interactions. Remarkably, we find that this lineshape structure is {to some extent }independent of the specific lattice geometry and detection and drive directions, but yet is sensitive to the ordered arrangement of atoms.  
The triplet lineshape therefore characterizes atomic arrays and sets them apart from  
non-interacting emitters and interacting emitters in disordered configurations \cite{jonesOlmos2017}. 
It could be measured in experiments \cite{ruiBloch2020,srakaewZeiher2023,olmosLesanovsky2013,coveyPainter2019}, where it would provide a clear experimental signature of strong coupling in non-linear regimes, beyond current demonstrations that are limited to linear phenomena \cite{ruiBloch2020,srakaewZeiher2023,glicensteinBrowaeys2020}.  

These predictions are made introducing a novel Dynamical Mean Field Theory (DMFT) approach to the steady-states of atomic arrays, and comparing with exact diagonalization calculations.
DMFT has been very successful in the context of  
strongly-correlated electrons \cite{georgesRozenberg1996,aokiWerner2014}, and describes the limit of large lattice dimensionality. 
{Here we leverage a DMFT approach to Markovian open quantum systems \cite{scarlatellaSchiro2021} and to spin systems in equilibrium \cite{lenkEckstein2022a}, to formulate a novel DMFT approach for atomic arrays. }
This paves the way to investigate a broad range of phenomena in these systems beyond a Gutzwiller mean-field approximation, including steady-state superradiance \cite{ostermannYelin2023}, lasing and interaction-induced transparency \cite{langPiazza2020a}. It complements other methods such as generalized mean-field approaches based on a factorization of spatial cumulants \cite{kramerRitsch2015, plankensteinerRitsch2022}, that break down when these describe important correlations \cite{robicheauxSuresh2021,kramerRitsch2015}, and a recent Truncated Wigner Approximation \cite{minkFleischhauer2023}, inaccurate when subradiant states are significant.

\begin{figure}
    \centering
    \includegraphics[width=0.9\linewidth]{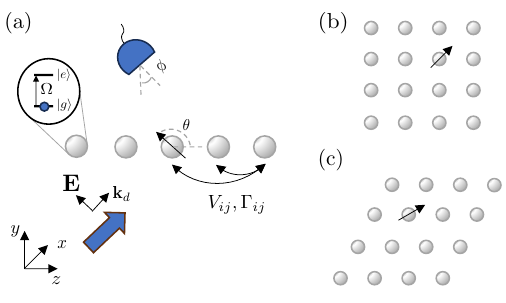}
    \caption{An array of two-level atoms driven by an electric field. The atoms are subject to coherent dipole-dipole interactions and collective dissipation. The photons emitted are collected by a detector. A 1D geometry is shown in (a),  a 2D square lattice in (b) and a triangular lattice in (c). The arrows indicate the orientation of the induced dipoles.
    }
    \label{fig:schematics}
\end{figure}


\textit{Model --} We consider a large number $N$ of two-level atoms ordered in a periodic array. The atoms are illuminated with a uniform plane wave and are coupled to the free-space electromagnetic field, giving rise both to coherent dipole-dipole interactions and to collective dissipation. 
These can be described by means of a Markovian master equation \cite{lehmbergLehmberg1970}: 
\begin{align}
\label{eq:atLightME}
\dot{\rho} &=-\frac{\ii}{\hbar}[H_{0} + H_{\rm drive}, \rho]+\mathcal{D}[\rho] \\
\label{eq:ham_0}
H_{0} &=  \frac{\hbar \Delta}{2} \sum_i^N {\sigma}_i^z + \sum_{i j ; i \neq j}^N  \hbar V_{i j} \sigma_i^{+} \sigma_j^{-} \\  
\label{eq:ham_drive}
H_{\rm drive} &= \frac{\hbar \Omega}{2} \sum_i^N \lp {\sigma}_i^+ e^{\ii \mathbf{k}_d \cdot \mathbf{r}_i } + {\sigma}_i^- e^{-\ii \mathbf{k}_d \cdot \mathbf{r}_i } \rp \\
\mathcal{D}[\rho]&=\frac{1}{2} \sum_{i, j} \Gamma_{i j}\left(2 \sigma_i^{-} \rho \sigma_j^{+}-\sigma_i^{+} \sigma_j^{-} \rho-\rho \sigma_i^{+} \sigma_j^{-}\right) .
\end{align}
Here $\sigma_l^\alpha$ with $\alpha=x,y$ or $z$ are the Pauli matrices on site $l$ and ${\sigma}_l^{ \pm}={\sigma}_l^x \pm \ii {\sigma}_l^y$ the raising and lowering operators; $\Delta = \omega_d - \omega_0$ is the detuning of the drive from the two-level transition energy $\omega_0$, $\Omega=2 \mathbf{d} \cdot \mathbf{E} / \hbar$ the Rabi coupling given by the vector of transition dipole moments $\mathbf{d}$ and the electric field vector $\mathbf{E}$, and $\mathbf{k}_d$ is the drive momentum.  

The dipolar interactions and collective decay terms are given by \cite{kramerRitsch2015,asenjo-garciaChang2017}
\begin{align} 
\label{eq:vij}
&V_{i l} =\frac{3 \Gamma}{4} \lbr-\alpha \frac{\cos (k_0 r_{ij})}{(k_0 r_{ij})}+\beta \lsq \frac{\sin (k_0 r_{ij})}{(k_0 r_{ij})^2}+\frac{\cos (k_0 r_{ij})}{(k_0 r_{ij})^3}\rsq \rbr \\
\label{eq:gamij}
&\Gamma_{i l} = \frac{3 \Gamma}{2} \lbr \alpha \frac{\sin (k_0 r_{ij})}{(k_0 r_{ij})}+\beta\lsq \frac{\cos (k_0 r_{ij})}{(k_0 r_{ij})^2}-\frac{\sin (k_0 r_{ij})}{(k_0 r_{ij})^3}\rsq \rbr  
\end{align}
where the single-atom decay rate is $\Gamma=$ $|\mathbf{d}|^2 k_0^3 / 3 \pi \epsilon_0 \hbar$ with  
$r_{i l}=\left|\mathbf{r}_i-\mathbf{r}_l\right|=a|i-l|$, $\mathbf{r}_i$ is the position of atom $i$ on the lattice and $a$ the lattice spacing. 
$k_0 = 2\pi/\lambda_0$ is the wavevector of two-level transition, with wavelength $\lambda_0$. $\alpha=1-\cos ^2 \theta$ and $\beta=1-3 \cos ^2 \theta$, where $\theta$ represents the angle between 
the common atomic dipole orientation and the line connecting atoms $i$ and $j$.


Throughout the manuscript, we will assume units of $\Gamma = 1$ and  $\hbar=1$. The interactions and losses are controlled by the parameter $k_0 a$: in the extreme limit $k_0 a \rw \infty$ interactions vanish and dissipation becomes local with $\Gamma_{il} = \Gamma \delta_{il}$, recovering the independent atoms case, while for $k_0 a \rw 0$ interactions diverge while dissipation becomes negligible, going to a finite value while also acquiring an all-to-all character $\Gamma_{il} = \Gamma$. Here we concentrate on intermediate values (see \cite{sm} for a plot).

\textit{DMFT for atomic arrays --} 
{We formulate a DMFT approach to investigate the steady-state of the model in the large atom number limit $N\rightarrow\infty$ (see \cite{sm} for details).  }
{To apply DMFT to the present case of spin Markovian master equation, we first map this onto a Keldysh action using spin-coherent states path integrals.} 
Then, we decouple the dipolar interactions and collective dissipation terms by introducing auxiliary bosonic fields, and apply a bosonic DMFT approach \cite{byczukVollhardt2008,huTong2009,andersWerner2010,andersWerner2011,strandWerner2015, scarlatellaSchiro2021}, on the lines of \cite{lenkEckstein2022a} and of an extended DMFT approach \cite{sunKotliar2002}.  
Due to the Markovian nature of the problem, we need to define a discretization of Green's functions at $t=0$ such that the Keldsyh causality structure \cite{kamenevKamenev2023} is preserved. 


Bosonic DMFT maps a lattice model onto an effective impurity model, describing a single site of the lattice coupled to an effective field, as in a standard Gutzwiller mean-field approach, and to an effective environment. Assuming a local self-energy in space, both can be determined such as to satisfy a set of self-consistent equations. {The effective environment introduces corrections of order $1/z$, where $z$ is the lattice coordination number, to Gutzwiller mean-field corresponding to $1/z \rw 0$ \cite{strandWerner2015}. In the present case in which interactions are not limited to nearest neighbors, the accuracy of DMFT is expected to improve. We later confirm our results by comparing with exact diagonalization (ED).}

In the present case, the impurity model is a generalized spin-boson model, that in the steady-state is described by the time-translation invariant Keldysh action 
\begin{equation}
\es{
\label{eq:n_sbAction}
S^{\mathrm{imp}}&=S_{0}^{i}+\oh  \int_{-\infty}^\infty d t \lsq \varsigma_{i}^\da(t) \tau_3 b  + \hc \rsq \\ 
&-\frac{1}{2} \int_{-\infty}^\infty d t \int_{-\infty}^\infty d t^{\prime} \varsigma_{i}^\da(t)  \tau_3 {\mathcal{W}}\left(t-t^{\prime}\right) \tau_3 \varsigma_{i}\left(t^{\prime}\right) 
}
\end{equation}
Here $\varsigma_i = (d_{i+}, \bar{d}_{i+},d_{i-}, \bar{d}_{i-} )^T$ is a vector in Nambu and Keldysh formalisms, {where $d=\aver{\sigma^-}$ and $\bar{d}=\aver{\sigma^+}$ are average values on spin coherent states \cite{altlandSimons2012}}, and the $+$ and $-$ indices indicate to which branch of the Keldysh double contour the fields belong to. $b$ and $\mathcal{W}$ represent the effective field and effective environment. 
These are taken site-independent assuming a homogeneous steady-state, as verified by a linear stability analysis. 
$\tau_3$ is a diagonal matrix with entries $(1,1,-1,-1)$ arising in the Keldysh-Nambu formalism.
Note that here a Nambu formalism is needed to allow for finite values of the anomalous correlators $\left\langle\sigma^{ \pm} \sigma^{ \pm}\right\rangle$, that arise due to the drive term, breaking the $U(1)$ symmetry of the undriven problem. 

The self-consistent conditions correspond to a set of matrix relations in frequency and momentum space: 
$\Pi_{\mathrm{loc}}=\left[ \tau_3{\chi}^\mo_{ii}\tau_3  + \mathcal{W}\right]^{-1} $, $U_{ii}=\frac{1}{N} \sum_{k} \left[W_{k}^\mo -\Pi_{\mathrm{loc}} \right]^{-1} $ and  $\mathcal{W}^{-1}=\Pi_{\mathrm{loc}}+U_{ii}^{-1}$ , and  $b_{\pm}=\left(  \mathcal{W}^R(\w={0})-  W_{k=0}^R(\w=0)\right) \aver{\varsigma_{i\pm}}$. Here the superscript ``$R$'' indicates the retarded component and $b_\pm$ and $\aver{\varsigma_{i\pm}}$ the $+$ or $-$ components of the corresponding vectors.  
$W_k$ is a known matrix with entries defined by the Fourier transforms of \eqref{eq:vij} and \eqref{eq:gamij}, reported in \cite{sm}. 
Assuming one can compute from the impurity action \eqref{eq:n_sbAction} the local connected Green's function $
\chi_{ii}(\tau)= -\ii \lim_{t\rw \infty} \left\langle  \varsigma_{i}(t+\tau) \varsigma^\da_{i}(t)\right\rangle_{S_{\text {imp }}^{\text {con }}}$
and expectation value $\left\langle\varsigma_{i}\right\rangle$, then these equations determine $\mathcal{W},b,\Pi_{\mathrm{loc}}$ and $U_{ii}$, such that the impurity problem represents the original lattice problem. 

The connected propagator of the lattice $\chi_{ij}(\tau)= -\ii \lim_{t\rw \infty} \left\langle  \varsigma_{i}(t+\tau) \varsigma^\da_{j}(t)\right\rangle^{\textrm{con}}$ can then be computed  by $\chi^\mo(k,\omega) = \tau_3 (\Pi_{\rm loc}^\mo(\omega) - W_k)\tau_3$.

Solving the impurity model is a challenge by itself and a viable numerically exact approach is still lacking.
Here we use a method based on a non-crossing approximation (NCA) \cite{schiroScarlatella2019,scarlatellaSchiro2021,scarlatellaSchiro2024a}. More details on DMFT/NCA are found in \cite{sm}.

\begin{figure}
    \centering
    \includegraphics[width=1\linewidth]{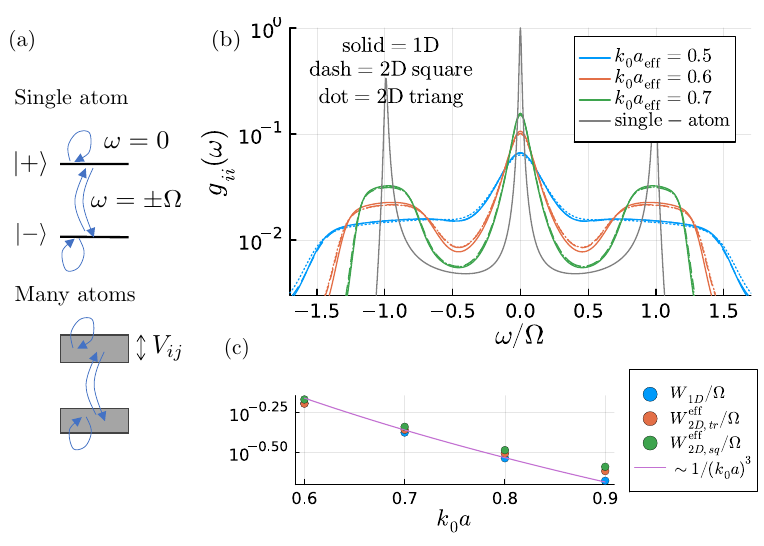}
    \caption{(a) Sketch of the single-particle transitions determining the Mollow triplet. In the case of many atoms (bottom) they take place between two bands
    determined by the dipolar interactions.
    (b) The local atomic correlation function for a 1D chain (solid lines) and 2D {square and triangular }lattices as defined in the body (dashed and dotted lines) as a function of lattice spacing and in DMFT, {for $N\rightarrow\infty$}. (c) The width of the sidebands $W$, extracted by the extrema of the derivatives, scales as the dipolar interaction \eqref{eq:vij} with a leading order $1/(k_0 a)^3$ behaviour.
  {The widths for the different geometries collapse on the 1D curve once rescaled as $W^{\rm eff} = W \Vert V_{1D} \Vert_2/\Vert V \Vert_2$.
Similarly, in (b) the 2D curves are plotted for a lattice spacing $a$ such that $\Vert V(a) \Vert_2  = \Vert V_{1D}(a_{\rm eff}) \Vert_2 $. 
 }
Here $\Omega=60\Gamma$, {dipoles point as in Fig. 1 (with $\theta=0$ in 1D) and drive is along $x$ (orthogonal to the array).} 
        }
    \label{fig:fig2}
\end{figure}


\textit{Mollow triplet in DMFT -- } 
The spectrum of the light emitted from a single two-level atom
which is strongly and continuously driven on resonance, $\Omega \gg \Gamma$ and $\Delta=0$, shows a central peak at the two-level transition energy $\omega = \omega_0$, accompanied two side peaks at $\omega = \omega_0 \pm \Omega$, known as the Mollow triplet \cite{mollowMollow1969,scullyZubairy1997,sibalicAdams2024}. 
This is understood by a picture in which the atom is dressed by the laser field \cite{cohen-tannoudjiReynaud1977,scullyZubairy1997}. 
In a frame rotating at the atomic frequency $\omega_0$, where energy differences are shifted by this quantity, the three resonances correspond to incoherent transitions induced by the jump operator $L=\sqrt{\Gamma}\sigma^-$ between the eigenstates $\ket{+}$,$\ket{-}$ of the single atom Hamiltonian $H=\Omega \sigma^x/2$, with transition energies $\omega=0$ and  $\omega=\pm \Omega$.  
These are represented graphically in Fig. \ref{fig:fig2} (a).

In the case of several atoms and for $V_{ij} \ll \Omega$, single-particle transitions can be understood as taking place between 
two bands of width set by $V_{ij}$, separated by the drive intensity $\Omega$, as depicted in Fig. \ref{fig:fig2}. 
The width of the bands corresponds to a broadening of the triplet lines, where interband transitions correspond to the central peak and intraband ones to the sidebands. 
Predicting how triplet broadens is nevertheless non-trivial. In fact, no broadening is predicted by a Gutzwiller mean-field approach, as well as by approximating the lattice self-energy with that of a single site, that instead allows to analytically estimate the Hubbard bands of a Mott insulator \cite{georgesRozenberg1996,strandWerner2015a}.  This motivates a DMFT approach.

The spectrum of emitted light in the case of atomic arrays is related to the atomic correlation functions $g_{ij}(\tau) = \lim_{t\rw \infty}\aver{\sigma_i^+(t+\tau) \sigma_j^-(t)}$ \cite{lehmbergLehmberg1970}. 
In DMFT, a key role is played by the local Green's function $g_{ii}(\tau)$. This could be measured by near-field microscopy \cite{courjonBainier1994,courjonCourjon2003} {and also characterizes the far-field emission, as discussed later. }
It is shown in Fig.~\ref{fig:fig2}~(b) for different values of $k_0 a$ and different geometries, namely a 1D chain with dipoles along the chain ($\theta=0$) and 2D square and triangular lattices with dipoles oriented as in Fig. \ref{fig:schematics}. The drive direction is chosen along $x$. 



We see that far from the limit of large $k_0 a$, where the single-atom triplet  composed of three Lorentzian peaks is recovered, the resonances acquire a characteristic lineshape with the sidebands becoming flat, while the central resonance remains peaked. They eventually merge when  $\Omega \lesssim V_{ij}$. 
Fig. \ref{fig:fig2} (c) shows that the resonance widths are set by the dipolar interactions, scaling like the nearest-neighbor contribution with a power of $1/(k_0 a)^3$ at small $k_0 a$.
This is in contrast with the case of a single atom, where the peaks widths are determined by the decay rate \cite{scullyZubairy1997}. 

Remarkably, the lineshape remains unchanged for all geometries {within our DMFT approach}.
We  find that it depends on the effective interaction parameter $\Vert V \Vert_2 = \sqrt{\sum_j V_{ij}^2}$  (see \cite{sm}). 
Curves for different geometries superimpose when plotted for a lattice spacing $a$ such that $\Vert V(a) \Vert_2  = \Vert V_{1D}(a_{\rm eff}) \Vert_2 $, where $a_{\rm eff}$ is a 1D-effective lattice spacing, as shown in Fig.~\ref{fig:fig2}~(b). 
Similarly, in Fig.~\ref{fig:fig2}~(c) we show that the sideband widths fall on the same curve when rescaled by this factor. 
Similar results are also found as a function of dipoles orientation in \cite{sm}. 
While the specific geometry is not important, a regular arrangement still is, as the characteristic flat sideband structure we find differs markedly from the structureless broadening expected in disordered ensembles \cite{jonesOlmos2017}.

A similar spectrum also characterizes the light emitted in the far field. 
This is proportional to the momentum-resolved propagator $g(\mathbf{k},\omega)=\int^{\infty}_{-\infty} d\tau \sum_{ij}   e^{\ii (\omega \tau + \mathbf{k} \cdot \mathbf{r}_{ij})} \lim_{t\rightarrow \infty} \left\langle\sigma_i^{+}\left(t + \tau \right) \sigma_j^{-}\left(t\right)\right\rangle$, computed at  $\mathbf{k}=k_0 {\mathbf{n}}$ where ${\mathbf{n}}$ is a unit vector indicating the emission direction \cite{lehmbergLehmberg1970}. {Note that only momenta $\abs{\mathbf{k}}\leq k_0$
are probed by the far-field emission, where the largest momentum is attained for an emission direction ${\mathbf{n}}$ parallel to the array. }
In DMFT, and for the same lattice geometries as in Fig. \ref{fig:fig2}, we find that this propagator becomes effectively momentum independent in the regime of large drive of the Mollow triplet and coincides with the local propagator $g(\mathbf{k},\omega) \approx g_{ii}(\omega)$.  {
The predicted lineshape therefore also characterizes the far-field emission in a generic emission direction, as we further discuss below.}

\textit{Validation with exact diagonalization -- } 
{A more complex dependence on geometry and momentum is expected beyond DMFT, which assumes a momentum-independent self-energy. Nevertheless, here we confirm 
that DMFT still captures correct qualitative features of the Mollow triplet broadening, 
comparing with exact diagonalization (ED) in the worst case scenario of a 1D chain, for which the largest deviations from DMFT (describing the large dimensional limit) are expected. }


For ED calculations we assume open boundary conditions.
To access larger systems sizes, we disregard the dissipator as the triplet broadening is mostly determined by dipolar interactions and these dominate at small  $k_0 a$ (see also \cite{sm}). 
Also, in this large-drive regime the stationary-state is found to be featureless and to approach an infinite temperature state. 

The local Green's function can then be approximated by weighting the density of states  (DoS) $\sum_{n,m} \delta({\omega - E_{n} - E_{m}})$ with the matrix elements $| \left[ \sigma_{i}^- \right]_{nm} |^2 = | \bra{\psi_n} \sigma_{i}^- \ket{\psi_m}|^2$,  see e.g. \cite{scarlatellaSchiro2019a}, namely $  g_{ii}(\omega) \approx 2 \pi/2^N \sum_{n,m} | \left[ \sigma_{i}^- \right]_{nm} |^2 \delta({\omega - E_{n} - E_{m}})$, where $\ket{\psi_n}$ and $E_n$ are the eigenstates and eigenvalues of the Hamiltonian. 
This is compared with the DMFT result in Fig.~\ref{fig:fig3}~(a), displaying very good agreement. 
We checked that the different broadening of the three peaks results from the matrix elements, probing the many-body eigenstates, while is not reflected in the DoS, as expected from the two-band picture sketched in Fig. \ref{fig:fig2}~(a).


\begin{figure}
    \centering
    \includegraphics[width=1\linewidth]{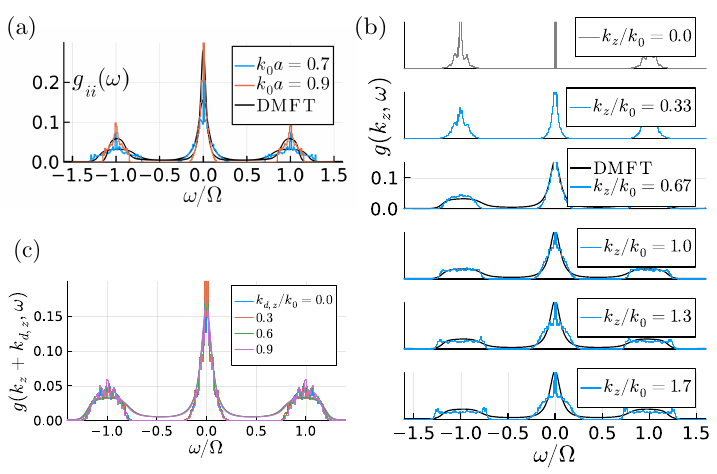}
    \caption{A comparison of the DMFT Green's functions {for $N\rightarrow\infty$} (smooth curves), with an approximation from exact diagonalization (ED) of the Hamlitonian for a chain of $N=10$ sites, calculated as a histogram (noisy curves). 
    In all panels $\Omega=60\Gamma$ and the drive is along $x$.
   Panel (a) shows the local Green's function.
   Panel (b) shows the momentum-resolved Green's function $g(k_z,\w)$ for $k_0a = 0.7$. 
   At $k_z=0$ the gray curve indicates that the ED-based approximation breaks down. 
   In these panels dipoles point along $z$ ($\theta=0$).
   Panel (c) shows the momentum-resolved Green's function for different drive directions labelled by $k_{z,d}$. Here $k_z a = k_0 a = 0.5$ and $\theta=\pi/2$.
   }
    \label{fig:fig3}
\end{figure}


Similarly, we approximate the momentum-resolved Green's function with a weighted sum of the DoS with matrix elements 
$| \left[ \sigma_{k}^- \right]_{nm} |^2/N = | \bra{\psi_n} \sigma_{k}^- \ket{\psi_m}|^2/N$ in Fig.~\ref{fig:fig3}~(b). 
Remarkably, this compares well with the momentum-independent DMFT prediction for most values of $k_z$, but also reveals some expected momentum dependence beyond DMFT, and beyond finite-size effects \cite{sm}. 
Comparison with DMFT deteriorates close to $k_z=0$, where the resonances narrow down up to a delta-function-like central peak due to a total spin conservation at $k_z=0$ (see \cite{sm}), and the ED approximation of neglecting the dissipator is not justified there.
We remark that the special case $k_z=0$ was previously studied in \cite{jonesOlmos2017} for a small 1D chain, while the characteristic lineshape for other emission directions was not predicted there.

{The success of DMFT in capturing the lineshape broadening of the Mollow triplet can be understood from the physics being effectively short-ranged, despite the long-range character of interactions and dissipation \eqref{eq:vij}, \eqref{eq:gamij}. 
In fact, we find identical results for a model with only {\it local} dissipation  \eqref{eq:gamij}, even though the non-local contributions are a priori non-negligible (see \cite{sm}). 
}


Finally, we recall that the far-field emission can only probe the Brillouin zone for $\abs{k_z}\leq k_0$, where $|k_z| = k_0$ corresponds to a direction of emission parallel to the chain ($\phi=\pm\pi/2$ in Fig. \ref{fig:schematics} (a)).
Nevertheless, there is a finite window of momenta in which such a broadening can be observed, as shown in Fig.  \ref{fig:fig3}~(b). {Changing the drive direction provides a natural way around this constraint.  }

\textit{Drive direction, atomic polarization -- }
{The triplet lineshape also does not depend significantly on specific choices of drive direction, as shown here, and atomic polarization discussed in \cite{sm}. } 
Fig. \ref{fig:fig3}~(c) shows the momentum resolved propagator at $k_z = k_0$ for different drive directions. 
These are accounted for within DMFT by a gauge transformation $\sigma_j^+ e^{i k_{d,z} a j}\rightarrow \sigma_j^+ $ that trades the drive phases with a shift in momentum of dipolar interactions and dissipation $V_{k_z} \rightarrow V_{k_z-k_{d,z}}$ and $\Gamma_{k_z} \rightarrow \Gamma_{k_z-k_{d,z}}$.
Both in DMFT and ED the Green's functions depend little on drive direction apart that they are shifted in momentum by $k_{d,z}$. This allows to scan the Brillouin zone by varying the drive direction, keeping the detection direction fixed.

{
\textit{Experimentally-relevant parameter regimes -- } 
The model studied finds natural experimental realizations \cite{ruiBloch2020,glicensteinBrowaeys2020,olmosLesanovsky2013,coveyPainter2019}, where optical lattices or tweezers are used to trap the atoms in ordered arrays (see also the discussion in \cite{ruostekoskiRuostekoski2023}). 
The broadened Mollow triplet is more pronounced for small lattice spacings (large peaks widths) and large drive strengths (large peaks separation), such as considered in previous figures, but these parameters are limited in experiments. 
The main experimental challenge is to reach deep subwavelength regimes, and the smallest lattice spacing of $k_0 a = 4.4$ was achieved experimentally in \cite{ruiBloch2020,srakaewZeiher2023}, using  ${ }^{87} \mathrm{Rb}$ atoms in a 2D square optical lattice. 
 
In Fig. \ref{fig:exper} we report DMFT calculations with parameters that are relevant for this experimental setup.
We note that the Mollow triplet broadening is already noticeable for the lattice spacing achieved in the experiment.  
The characteristic lineshape will be instead better resolved in the next generation of experiments that will realize shorter lattice spacings, such as those proposed in \cite{olmosLesanovsky2013}  and \cite{coveyPainter2019}. 
}



\begin{figure}
    \centering
        \includegraphics[width=0.65\linewidth]{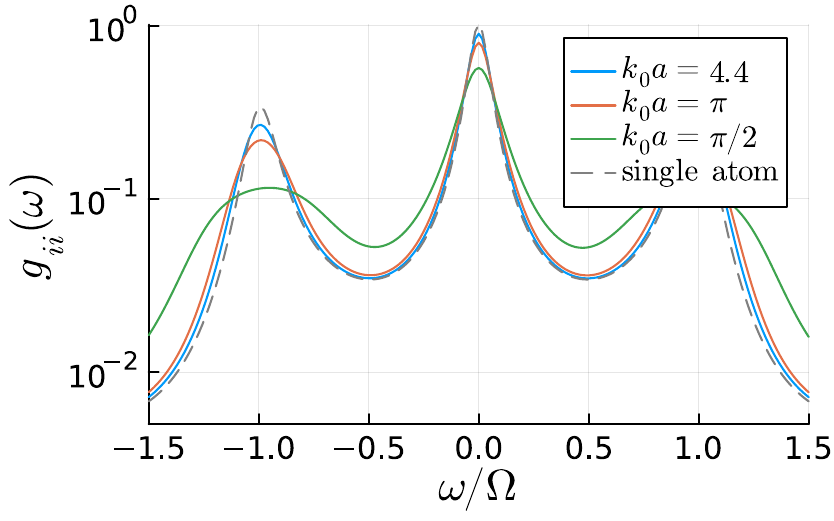}
    \caption{The local atomic correlation function for experimentally relevant parameters for the realization in \cite{ruiBloch2020}. 
This is obtained for a 2D square lattice geometry and within DMFT for $N\rightarrow\infty$. Here $\Omega/\Gamma=10$, dipoles point along $z$ and the drive is along $x$ (orthogonal to the array).
}
    \label{fig:exper}
\end{figure}


\textit{Conclusions -- } We showed that the spectrum of light emitted by strongly-driven atomic arrays is characterized by a modified Mollow triplet whose lineshape is determined by dipolar interactions, {to some extent }independent of lattice geometry, emission and drive directions. 
This sets atomic arrays apart from disordered ensembles and non-interacting emitters and would provide a clear experimental signature of strong coupling in non-linear regimes.

We have also introduced a Dynamical Mean-Field Theory approach for atomic arrays, paving the way for investigating a variety of strongly-correlated phenomena in these systems. 
Our steady-state approach could be extended to the transient dynamics \cite{aokiWerner2014}, where superradiance or the generation of subradiant states could be investigated \cite{rubies-bigordaYelin2023}.

\textit{Acknowledgements -- } We thank Mike Gunn for insightful discussions. 
This work was supported by the Engineering and Physical Sciences Research Council [grant number EP/W005484], and a Simons Investigator Award [Grant No. 511029].
For the purpose of open access, the authors has applied a creative commons attribution (CC BY) licence to any author accepted manuscript version arising.

The data (code) to reproduce the results of the manuscript will provided by the author under request.

\putbib
\end{bibunit}

\bibliography{}

\newpage $ $ \newpage
\appendix
\onecolumngrid
\begin{bibunit}

 \title{Supplemental Material to ``Fate of the Mollow triplet in strongly-coupled atomic arrays''}
 \author{Orazio Scarlatella}
 \author{Nigel R. Cooper}
 \affiliation{T.C.M. Group, Cavendish Laboratory, University of Cambridge, J.J. Thomson Avenue, Cambridge CB3 0HE, UK\looseness=-1}
 \maketitle
 \onecolumngrid

\setcounter{equation}{0}
\setcounter{page}{1}
\setcounter{figure}{0}
\renewcommand{\thefigure}{S.\arabic{figure}}


This supplemental is organized as follows. 
In App.  \ref{app:model} we give additional details on the model,  
in App. \ref{app:tipVsDiss} we discuss further the triplet dipendence on parameters. In App. \ref{app:finSizeED} we report a further discussion of exact diagonalization results. 
The DMFT formalism is discussed in App. \ref{app:dmftForm}, its numerical implementation in App. \ref{sec:numImp}, the NCA impurity solver in App. \ref{app:nca} and the definition of the Green's functions at $t=0$ in App. \ref{sec:t=0Greens}.

\section{Details of the model}
\label{app:model}

\begin{figure}[h]
    \centering
    \includegraphics[width=0.6\linewidth]{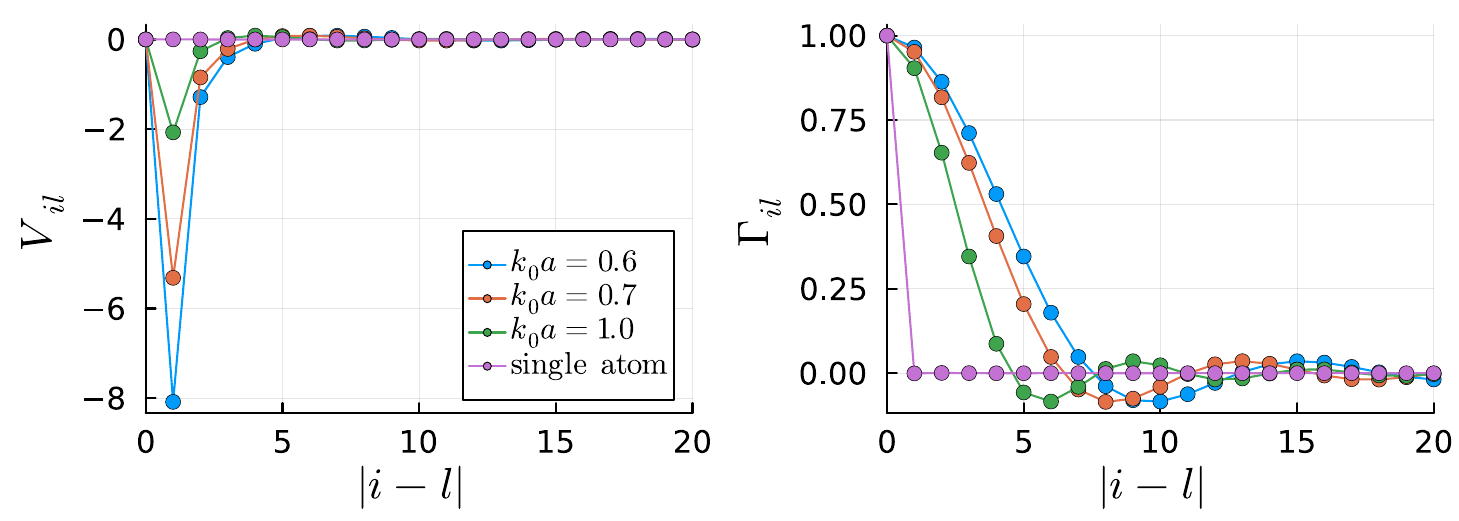}
    \caption{Dipolar interactions $V_{il}$ and collective dissipation rate $\Gamma_{il}$ for several values of $k_0 a$  for a 1D chain and for $\theta = 0$.}
    \label{fig:gamV}
\end{figure}

The dipolar interaction and collective dissipation are plotted in Fig. \ref{fig:gamV} for several values of $k_0 a$ and in 1D. Note that for small values of $k_0 a$ the interaction becomes larger than the dissipation rate.

\section{Other results in DMFT}
\label{app:tipVsDiss}

\begin{figure}
    \centering
    \includegraphics[width=0.5\linewidth]{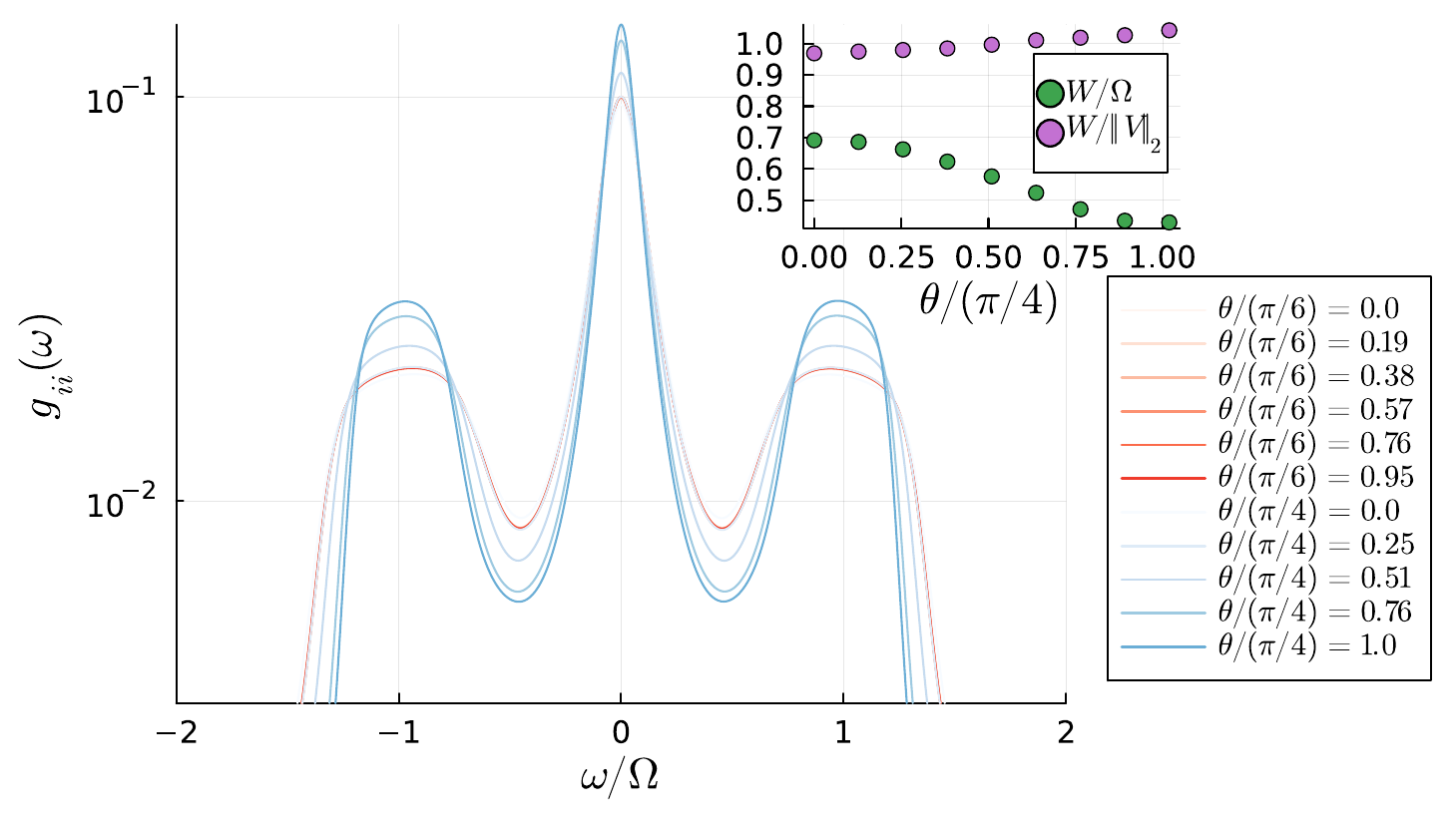}
    \caption{The local atomic Green function for a square (blue) and triangular (red) lattice, as a function of the dipole orientation, for in-plane diples and at an angle $\theta$ with the $z$ direction. The inset shows that the sideband widths depend mainly on $\Vert V \Vert_2 = \sqrt{\sum_j V_{ij}^2}$ in the case of the square lattice.  }
    \label{fig:2D_polarization}
\end{figure}
{In the main text we discussed that the lineshape in DMFT is found to be independent of the specific geometry and to depend on an effective interaction parameter $\Vert V \Vert_2 = \sqrt{\sum_j V_{ij}^2}$. 
This is found numerically based on the following analysis.  
Fig. \ref{fig:2D_polarization} shows that changing the in-plane dipole orientation essentially does not affect the lineshape for a triangular lattice, while does so for a square lattice. In the first case, the parameter $\Vert V \Vert_2 $ is strictly constant for the nearest neighbors dipolar interactions, and almost constant beyond nearest neighbors. 
For a square lattice we find that the varying widths roughly depend on $\Vert V \Vert_2 $, as shown in the inset. In Fig. \ref{fig:fig2} panel (c) we showed that the same dependence is found at fixed dipole orientations and across the different geometries (1D, 2D square lattice and triangular lattice).  
}

\begin{figure}[h]
    \centering
    \includegraphics[width=0.55\linewidth]{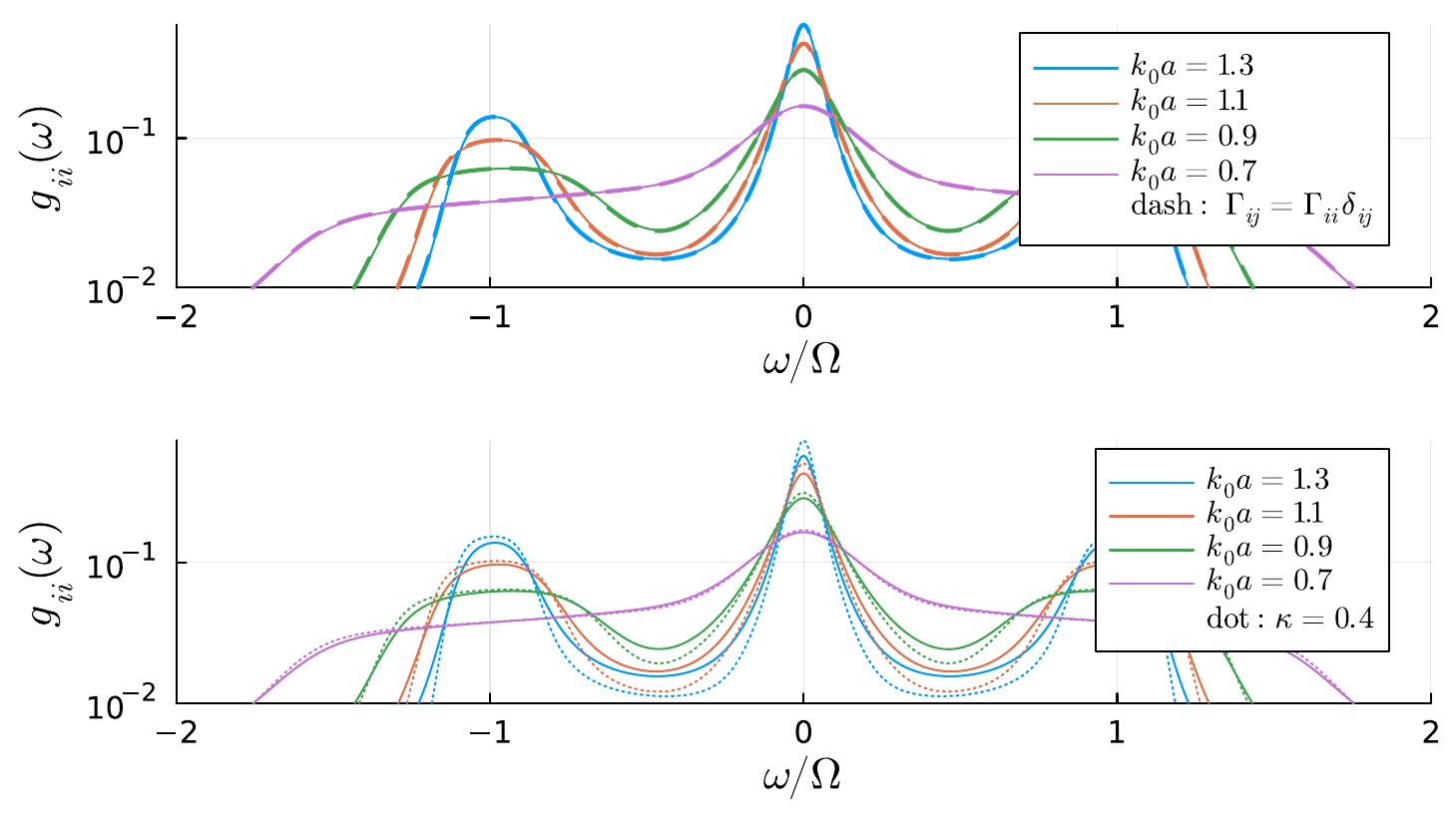}
    \caption{Local atomic correlation function in DMFT for a 1D chain, for $V_{ij},\Gamma_{ij}$ as defined in the main text (solid), compared in the top panel to cases in which the dissipation is made local $\Gamma_{ij}\rw\delta_{ij}\Gamma_{ii}$ (dashed line). 
    In the bottom panel the solid line is compared with the case in which dissipation strength is changed $\Gamma_{ij} \rw \kappa \Gamma_{ij}$ keeping the dipolar interactions fixed (dotted line).     
    Here $\Omega=20\Gamma$ , dipoles point along $z$ and drive along $x$ (orthogonal to the array).}
    \label{fig:tripVsDiss}
\end{figure}

In the top panel of Fig. \ref{fig:tripVsDiss}  we show that switching to a local dissipation $\Gamma_{ij}\rw\delta_{ij}\Gamma_{ii}$ leaves the DMFT prediction for the triplet unchanged. 
We notice though that the non-local contribution to the dissipation  \eqref{eq:gamij} is not parametrically negligible with respect to the local one (see Fig. \ref{fig:gamV}). 
This shows the triplet broadening effectively results only from short-ranged processes.
In the bottom panel of Fig. \ref{fig:tripVsDiss} we show that by artificially changing the dissipation strength $\Gamma_{ij} \rw \kappa \Gamma_{ij}$ while keeping the dipole interactions constant essentially leaves the sidebands widths unchanged and mainly affects the amplitudes. 
This confirms that the triplet broadening is mostly determined by dipolar interactions rather than by the decay rate, and justifies neglecting the dissipator in exact diagonalization calculations.



\section{Other results in exact diagonalization}
\label{app:finSizeED}

\begin{figure}
    \centering
    \includegraphics[width=0.8\linewidth]{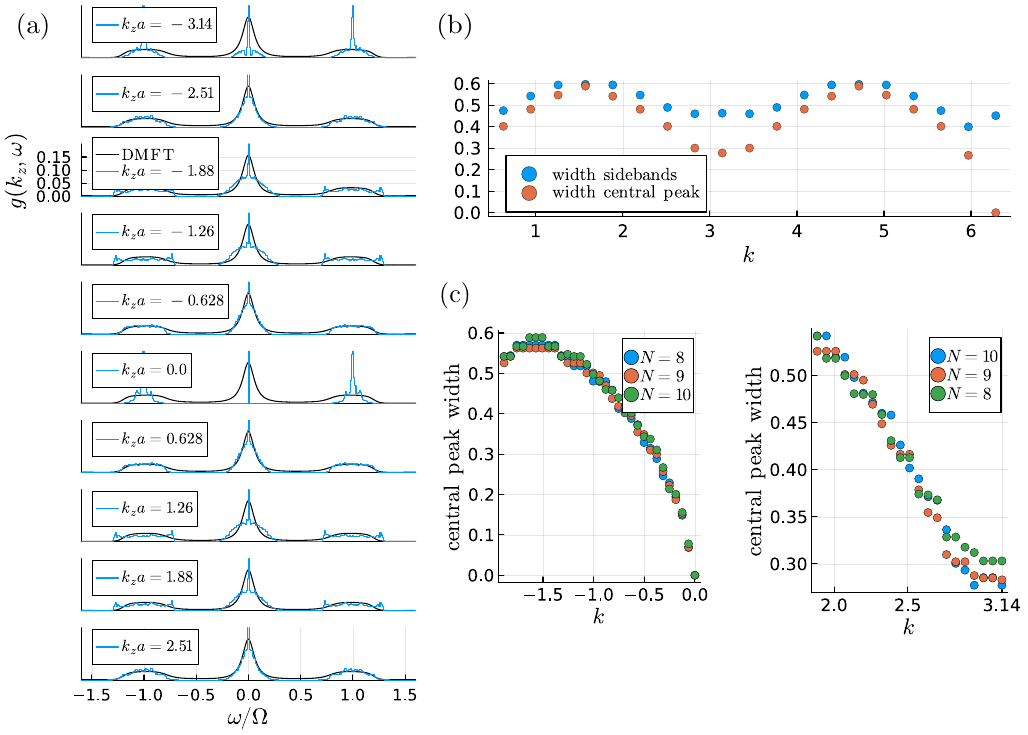}
    \caption{(a) Momentum-resolved propagator in DMFT and in ED, in the same approximation of neglecting the dissipator as in the main text, scanning the whole Brillouin zone. 
    (b) and (c): the widths of the sidebands and central peak vs momentum, extracted from the momentum-resolved matrix element, defined by a threshold value on the amplitude of $0.05$. (a),(b) are for N=10 atoms. (c) shows the dependence on system size $N$, respectively close to $k=0$ and $k=\pi$.}
    \label{fig:scaling}
\end{figure}

Fig. \ref{fig:scaling} (a) shows the momentum-resolved propagator in ED, approximated as in main text, throughout the entire Brillouin zone. 
In Fig. \ref{fig:scaling} (b) and (c) the widths of the triplet resonances are extracted from the momentum-resolved matrix element in ED by a threshold value on the amplitude.
Figures~\ref{fig:scaling} (a) and (b) show that the momentum dependence is most pronounced around the points $k_z a=0$ and $k_z a=\pi$ where the widths of the resonances decreases significantly and comparison with DMFT is the worst.
Fig.~\ref{fig:scaling} (c) shows that the same behaviour is found changing the system sizes, suggesting that those features are not finite-size effects, but rather a true momentum dependence that cannot be resolved by DMFT. 

\subsubsection{Narrow central peak at k=0}

The fact that the matrix element of $\sigma_{k=0}^- = \sum_i \sigma_i^- $ has a vanishing width for the central peak in Figs. \ref{fig:fig3} and \ref{fig:scaling} arises due to total spin conservation. 
This can be understood in perturbation theory. For this we need to calculate the matrix element of $\sum_i \sigma_i^- $ on the eigenstates of the Hamiltonian. 
It is useful to make a rotation such that $\sigma^x \rw \sigma^z$, for which $\sigma^- = (\sigma^x - i \sigma^y)/2 \rw (\sigma^z - i \sigma^x)/2$ and the Hamiltonian reads $H \rightarrow \frac{\Omega}{2}\sum_i \sigma_i^z + \sum_{ij} V_{ij}/4 (\sigma_i^z + i \sigma_i^x) (\sigma_i^z - i \sigma_i^x) $. 
Assuming $\Omega\gg V$ and treating $V$ as a perturbation, the perturbed eigenstates at zeroth order are labelled by $\ket{\sigma_{\rm tot}^z,n}$, obtained by diagonalizing the perturbation in the degenerate subspaces of the non-interacting Hamiltonian at fixed $\sigma_{\rm tot}^z$. 
In the new basis, we want to compute $\bra{\sigma_{\rm tot 1}^z,n_1} \sum_i (\sigma_i^z - i \sigma_i^x)/2 \ket{\sigma_{\rm tot2}^z,n_2}$, where  
the transitions contributing to the central peak are given by the case $\sigma_{\rm tot 1}^z=\sigma_{\rm tot 2}^z$. In this case $\bra{\sigma_{\rm tot }^z,n_1} \sigma_i^x \ket{\sigma_{\rm tot}^z,n_2}=0$ because $\sigma_i^x = \sigma_i^+ + \sigma_i^-$ changes total $\sigma^z$ by 1 (expanding in the basis of tensor products of $\sigma^z_i$ eigenstates the overlap vanishes term by term), while $\sum_i\sigma_i^z$ gets replaced by its eigenvalue and the matrix element vanishes for $n_1 \neq n_2$ by orthogonality of different eigenstates
\begin{equation}
\bra{\sigma_{\rm tot}^z,n_1} \sum_i (\sigma_i^z - i \sigma_i^x)/2 \ket{\sigma_{\rm tot}^z,n_2} =\sigma_{\rm tot}^z \delta_{n_1,n_2}/2 
\end{equation}
This corresponds to the narrow central peak at $k=0$. The numerics show that this holds beyond this simple perturbative argument.
For $k \ll 1$ the pointed shape of the central peak can be understood as those $n_1 \neq n_2$ matrix elements, corresponding to a broadening of the central peak, acquiring a finite yet small weight compared to the $n_1=n_2$ terms. 

\subsubsection{Few atoms case}

\begin{figure}
    \centering
    \includegraphics[width=0.7\linewidth]{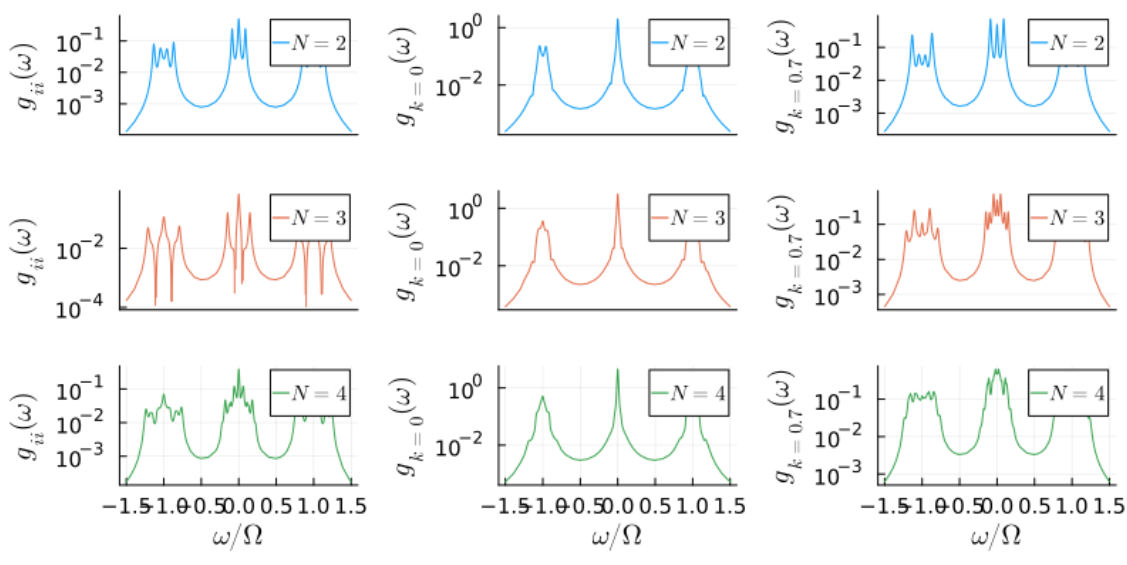}
    \caption{Green's function computed in ED by diagonalizing the full Liouvillian (including the dissipator) for increasing number of atoms and for $k_0 a = 0.7$, $\Omega=60$ and for $i=N/2$ ($i=(N+1)/2$) for even (odd) $N$. }
    \label{fig:fewAtoms}
\end{figure}

In Fig. \ref{fig:fewAtoms} we show the local propagator computed from an exact diagonalization of the full Liouvillian (including the dissipator). 
This shows how the broadening in the thermodynamic limit emerges from the limit of a few atoms. 
We note that in the case of very few atoms that can be solved analytically, is difficult to infer the structure in the thermodynamic limit.   

\section{Nonequilibrium steady-state DMFT}
\label{app:dmftForm}

Here we present the main steps in the derivation of the DMFT approach for the steady-state of the master equation \eqref{eq:atLightME}.  

\subsection{From master equation to Keldysh action for spin systems}

{The first step is to obtain the Keldsyh action corresponding to the Markovian master equation of spin operators \eqref{eq:atLightME}. 
For bosonic systems there's a standard procedure to do this for normal-ordered Hamiltonians and jump operators, using the bosonic coherent states path integral \cite{siebererDiehl2016,siebererDiehl2014}. 
For spin systems one needs to resort to the less familiar spin coherent states path integral \cite{altlandSimons2012}. In the present case, the action is given by 
$
S_{\text {eff }}=S_{0}+S_{\text {int }}
$,  
where the non-local terms are 
\begin{equation}
\es{
S_{\mathrm{int}} &=- \int_{-\infty}^\infty d t \left[ \sum_{ij, i \neq j}  V_{i j} \lp  \bar{d}_{i_+} d_{j+} - \bar{d}_{i-} d_{j-} \rp \right. \\ 
&+ \left. \frac{i}{2} \sum_{i, j} \Gamma_{i j}\left(2 d_{i+}  \bar{d}_{j-} - \bar{d}_{i+} d_{j+} - \bar{d}_{i-} d_{j-}\right) \right]
}
\end{equation}
while we don't need to specify the local action $S_0$ for our purposes.
$S_{\text {int }}$ takes the same form familiar from bosons, where $d$ is an average value on spin coherent states \cite{altlandSimons2012} ($\bar{d}$ its complex conjugate), rather than on bosonic coherent states. 
This action is derived starting from a non-Markovian, Caldeira-Leggett description of the bath in terms of non-interacting bosonic modes linearly coupled to the spin system, and then by integrating out the bath and making a Markovian approximation, as described for bosonic systems in \cite{siebererDiehl2014}. 
We note that this procedure yields a quadratic action in $d$ and $\bar{d}$. Instead a different result would be obtained if the path integral were constructed directly from the master equation \eqref{eq:atLightME}, inserting resolutions of identities with a similar procedure as described in \cite{siebererDiehl2016} for bosonic systems. In that case both the averages $\aver{L} = \aver{{\sigma}^-}$ and $\aver{L^\dagger L} = \aver{{\sigma}^+{\sigma}^-}$ taken on coherent states would appear in the action, but $\aver{{\sigma}^+{\sigma}^-} \neq  \aver{{\sigma}^+} \aver{{\sigma}^-} $ as opposed to the bosonic case, and this wouldn't be a quadratic form in $\aver{{\sigma}^-}$ and $\aver{{\sigma}^+}$ (thus in $d$ and $\bar{d}$). The difference between the two cases is that in the second (starting from the master equation) the path integral timestep is larger than the bath memory time, which is sent to zero first with a Markovian approximation, while the opposite is true in the first case of a Caldeira-Leggett bath. In the bosonic case the two procedures yield the same results, except when tadpole diagrams are computed \cite{siebererDiehl2014}. 
We also note that for the same reason a master equation with jump operators that are a normal ordered product of ${\sigma}^+$ and ${\sigma}^-$ operators (except the simple case above in which they are linear) does not map onto an action with a similar form to its bosonic counterpart  \cite{siebererDiehl2016}. }



For DMFT, it is convenient to use the Keldysh-Nambu vectors notation introduced in the main text. 
To work with unrestricted sums, we define $V_{ii} = 0$, and to use the Nambu formalism we note that $(2,2)$ Nambu component is related to the $(1,1)$ by transposing space-time and Keldysh indices. Note also that $\Gamma, V$ are symmetric $\Gamma_{ji}= \Gamma_{ij}$, $V_{ji}= V_{ij}$. The action then can be written as
\begin{equation}
\label{eq:intAction_realspace}
\es{
S_{\mathrm{int}}= - \oh \int_{-\infty}^\infty d t \sum_{ij} 
\bma \vdots \ema_i^\da 
\bma   {V}_{ij} - \frac{i}{2} {\Gamma_{ij}} & 0 & 0 & 0 \\ 
0 &    {V}_{ij} - \frac{i}{2} {\Gamma_{ij}} & 0 & i \Gamma_{ij} \\ 
i  {\Gamma_{ij}} & 0 & -   {V}_{ij} - \frac{i}{2} {\Gamma_{ij}} & 0 \\
0 & 0 & 0 & -  {V}_{ij} - \frac{i}{2} {\Gamma_{ij}}
\ema 
\bma 
 d_+ \\ \bar{d}_+  \\ 
 d_- \\ \bar{d}_- 
\ema_j
}
\end{equation}

In frequency and momentum space it becomes  
\begin{equation}
\es{
S_{\mathrm{int}} &= - \oh \int_{-\infty}^\infty \frac{d\w}{2\pi} \frac{1}{N}\sum_{k} 
\bma \vdots \ema^\da 
\bma   {V}_{k} - \frac{i}{2} {\Gamma_{k}} & 0 & 0 & 0 \\ 
0 &    {V}_{k} - \frac{i}{2} {\Gamma_{k}} & 0 & i \Gamma_{k} \\ 
i  {\Gamma_{k}} & 0 & -   {V}_{k} - \frac{i}{2} {\Gamma_{k}} & 0 \\
0 & 0 & 0 &  -  {V}_{k} - \frac{i}{2} {\Gamma_{k}}
\ema 
\bma 
d_+(\w,k) \\ \bar{d}_+ (-\w,-k)  \\ 
d_-(\w,k) \\ \bar{d}_-(-\w,-k) 
\ema  \\ 
&= - \oh \int_{-\infty}^\infty \frac{d\w}{2\pi} \frac{1}{N}\sum_{k} 
\bma \vdots \ema^\da 
\tau_3 W_k \tau_3
\bma 
d_+(\w,k) \\ \bar{d}_+ (-\w,-k)  \\ 
d_-(\w,k) \\ \bar{d}_-(-\w,-k) 
\ema  
}
\end{equation}
which defines the matrix $W_k$ appearing in the main text.


To proceed, we write the real space and real time action \eqref{eq:intAction_realspace} in the compact form 
\begin{equation}
\label{eq:n_lmSEn}
S_{\mathrm{int}}= - \frac{1}{2} \int_{-\infty}^\infty d t d t^{\prime} \sum_{r r^{\prime}} \varsigma_r^\da (t)\tau_3 W_{r r^{\prime}}\left(t-t^{\prime}\right) \tau_3 \varsigma_{r^{\prime}}\left(t^{\prime}\right)
\end{equation}
where we defined the site positions $r = i a$ and where in the present case of a Markovian master equation $W_{r r^{\prime}}\left(t-t^{\prime}\right) = W_{r r^{\prime}} \delta(t - t') $ is time local. 

In the following, we generalize the equilibrium DMFT approach of \cite{lenkEckstein2022a} to Keldysh field theory in the stationary state. Our derivation mostly follows this reference, therefore we report the main equations here and refer to  \cite{lenkEckstein2022a} for more details on the derivation.

\subsection{Bosonic representation of dipolar interaction and collective dissipation}

We introduce auxiliary bosonic fields $\phi_{r}(t)$ at each site using a Hubbard-Stratonovich transformation. Namely we define a new action $S_{\text {HS }}$ such that the original action is reproduced if the auxiliary gaussian fields are integrated out: $ e^{i S_{\text {eff }}}=\int \mathcal{D}[\phi] e^{i S_{\mathrm{HS}}}.$
Up to an irrelevant constant that we will omit throughout the manuscript, the Hubbard-Stratonovich action reads
\begin{equation}
S_{\mathrm{HS}}=S_{0}+S_{\phi \phi}+S_{\phi \sigma},
\end{equation}
with a quadratic term
$$
S_{\phi \phi}= \oh \int_{-\infty}^\infty  d t  d t^{\prime} \sum_{r r^{\prime}} \phi_{r}^\da(t)\tau_3 \left[W^{-1}\right]_{r r^{\prime}}\left(t-t^{\prime}\right) \tau_3 \phi_{r^{\prime}}\left(t^{\prime}\right),
$$
and a local linear coupling term
$$
S_{\phi \sigma}= \oh \sum_{r} \int_{-\infty}^\infty d t \phi^\da_{r}(t) \tau_3 \varsigma_{r}(t) + \hc
$$
where now $\phi$ is a Nambu vector of complex bosonic fields. 

\subsection{Impurity model and self-consistent equations}
The Hubbard-Stratonovic action defines a bosonic field theory coupled to spin fields, that introduce an anharmonicity for the bosons.  Since the coupling is local, we can apply a bosonic DMFT approach. 
This maps the Hubbard-Stratonovic action to the local impurity problem
\begin{equation}
\label{eq:n_bosAct}
\begin{aligned}
&S_{\mathrm{HS}}^{\mathrm{imp}}=S_{0}^{i}+S_{\phi \phi}^{\mathrm{imp}}+S_{\phi \sigma}^{\mathrm{imp}}, \\
&S_{\phi \sigma}^{\mathrm{imp}}= {+} \oh  \int_{-\infty}^\infty d t \phi^\da_{i}(t) \tau_3 \left[\varsigma_{i}(t)-h(t)\right] + \hc , \\
&S_{\phi \phi}^{\mathrm{imp}}=\frac{1}{2} \int_{-\infty}^\infty d t d t^{\prime} \phi^\da_{i}(t) \tau_3 \mathcal{W}^{-1}\left(t-t^{\prime}\right)\tau_3 \phi_{i}\left(t^{\prime}\right) .
\end{aligned}
\end{equation}
where $S_{0}^{i}$ is the action of a single isolated atom on site $i$. The interaction of the impurity site with the rest of the lattice is mimicked by the effective field $h(t)$, analogously an in a Gutzwiller mean-field approach, and by the effective Weiss field $\mathcal{W}$, assumed homogeneous in space. 

From the impurity model, one can compute the local bosonic expectation value $\left\langle\phi_{i}(t)\right\rangle_{S_{\mathrm{HS}}}$ and the connected correlation function 
$
U_{ii}(t-t')= -i \left\langle \phi_{i}(t) \phi^\da_{i}(t')\right\rangle_{S_{\mathrm{HS}}^{\mathrm{imp}}}^{\mathrm{con}}
$. 
This defines the impurity self-energy via the Dyson equation
\begin{equation}
\label{eq:n_dmft1}
\Pi_{\mathrm{loc}}=\mathcal{W}^{-1}-U_{ii}^{-1} .
\end{equation}
At the same time, assuming translation invariance, the lattice Green's function for the bosons
$
U_{r r^{\prime}}(t-t')=-i \left\langle  \phi_{r}(t) \phi^\da_{r^{\prime}}(t')\right\rangle_{S_{\mathrm{HS}}}^{\text {con }}
$ obeys in momentum space the Dyson equation
$
U_{k}^\mo=W_k^\mo - \Pi_k
$.  
In DMFT the self energy  $\Pi_k$ is assumed to be local in space and is identified with the impurity self-energy $\Pi_{k} = \Pi_{\text {loc }}$. Thus, summing over momentum, the local correlation function must also satisfy
\begin{equation}
\label{eq:n_dmft2}
\es{
U_{ii}&=\frac{1}{N} \sum_{k} \left[W_{k}^\mo -\Pi_{\mathrm{loc}} \right]^{-1} 
}
\end{equation}
which from \eqref{eq:n_dmft1} determines the Weiss field $\mathcal{W}$.  
In addition, assuming $\left\langle\phi_{r}(t)\right\rangle_{S_{\mathrm{HS}}}=\phi$ for all $r$ and $t$, the field $h$ is given by
\begin{equation}
\label{eq:dmft_2}
    h_{+(-)}= - \lp   [W_{k=0}^{-1}(\w=0)]^R-  [\mathcal{W}(\w=0)^{-1}]^R \rp \phi_{+(-)},
\end{equation}
where $h$ and $\phi$ take the same value on the $+$ and $-$ contours, as a classical field and an average value, and $R$ indicates the retarded component of the Fourier transform of the correlation function (this is a $2\times 2$ matrix in Nambu space).

Given an effective field $h$ and a Weiss field $\mathcal{W}$, the values of the local correlation function $U_{ii}$ and average bosonic field $\phi$ are completely determined by solving the impurity problem, with an appropriate method. But these values must also satisfy \eqref{eq:n_dmft1} \eqref{eq:n_dmft2}  and \eqref{eq:dmft_2}, for the impurity problem to be representative of the lattice model. These conditions therefore provide a set of closed equations for $h$ and $\mathcal{W}$, that must be satisfied by a DMFT solution.

\subsection{Alternative impurity model}

The impurity action \eqref{eq:n_bosAct} defines a spin-boson model, where a two-level system is coupled to a continuum of bosonic modes, with propagator $\mathcal{W}$. To solve the impurity model, it is convenient to integrate out the bosonic fields, leading to an equivalent spin model with action
\begin{equation}
\label{eq:n_sbAction_sm}
S^{\mathrm{imp}}=S_{0}^{i}+S_{\mathrm{int}, 1}^{\mathrm{imp}}+S_{\mathrm{int}, 2}^{\mathrm{imp}}
\end{equation}
with an effective field term
$$
S_{\mathrm{int}, 1}^{\mathrm{imp}}= \oh   \int_{-\infty}^\infty  d t \varsigma_{i}^\da(t) \tau_3 b  + \hc,
$$
and retarded interaction
$$
S_{\mathrm{int}, 2}^{\mathrm{imp}}=-\frac{1}{2}  \int_{-\infty}^\infty d t d t^{\prime} \varsigma_{i}^\da(t) \tau_3 {\mathcal{W}}\left(t-t^{\prime}\right) \tau_3 \varsigma_{i}\left(t^{\prime}\right) .
$$
Defining the spin connected Green's function as $
\chi_{ii}(t-t')= -i \left\langle \varsigma_{i}(t) \varsigma^\da_{i}(t')\right\rangle_{S_{\text {imp }}^{\text {con }}}$, this is related to the bosonic Green's function by 
$\quad U_{ii}=\mathcal{W}+\mathcal{W} \circ\tau_3 {\chi}_{ii} \tau_3 \circ \mathcal{W} $, where $\circ$ indicates real-time convolutions. 
Using Eq. \eqref{eq:n_dmft1}, this yields the expression for the self-energy in terms of the spin Green function 
\begin{equation}
\label{eq:n_dmft3}
\Pi_{\mathrm{loc}}=\left[ \tau_3 {\chi}^\mo_{ii} \tau_3  + \mathcal{W}\right]^{-1} 
\end{equation}
Also, from the gaussian integration, the effective field has expression $b_{+(-)}= \mathcal{W}^R(\w=0) h_{+(-)} $ and demanding that the action is stationary around the mean-field order parameter $\phi_{+(-)}$ one obtains $\left\langle\varsigma_{+(-)}\right\rangle$ as a function of $h_{+(-)}$. Combining those expressions yields
\begin{equation}
\label{eq:n_dmft4}
b_{+(-)}=\left(  \mathcal{W}^R(\w={0})-  W_{k=0}^R(\w=0)\right)\left\langle\varsigma_{i,+(-)}\right\rangle,
\end{equation}
These two equations together with \eqref{eq:n_dmft1} and \eqref{eq:n_dmft2} defines an equivalent set of DMFT equations to those given in the previous paragraph. These are in terms of spin quantities which are most natural to compute from our impurity solver.

\section{Numerical implementation of DMFT}
\label{sec:numImp}

Starting from a guess for $b$ and  $\mathcal{W}$, such as the mean field solution for the first and $\mathcal{W} = 1/N \sum_k W_{k}$ for the second, the following steps are iterated until a fixed point is reached: 

\begin{itemize}
\item \textit{Solving the impurity model}: given $\mathcal{W}$, $b$, the spin model is solved using the impurity solver, computing the steady-state density matrix $\rho_s$ and atomic correlation function 
$\chi = \bma \chi^{{++}}& \chi^{+-} \\ \chi^{-+} & \chi^{--}  \ema$ 
for $t>0$, where the entries are $2\times 2$ matrices in Nambu space. 
Then $\chi$ at negative times $t<0$ is obtained assuming the steady-state relation 
\begin{equation} 
\label{eq:ssCorrFuncs}
\chi_{\alpha \beta}^{ab}(-t) = - [\chi^{ba}_{\bar{\beta}\bar{\alpha}}]^* (t)
\end{equation}
where the conjugate-transpose of both Nambu $a,b$ and Keldysh $\alpha \beta$ indices is taken and the Keldysh indices $\alpha,\beta \in [+,-]$ are negated, such that $\bar{\alpha} = -\alpha$.
\item \textit{updating the Weiss field $\mathcal{W}$}: in frequency and momentum space, where inverses of Green's functions are simply given by a matrix inverse of their $4 \times 4$ Nambu-Keldysh stucture, we evaluate the self-consistent equations reported in the main text: $\Pi_{\mathrm{loc}}=\left[ \tau_3{\chi}^\mo_{ii}\tau_3  + \mathcal{W}\right]^{-1} $, $U_{ii}=\frac{1}{N} \sum_{k} \left[W_{k}^\mo -\Pi_{\mathrm{loc}} \right]^{-1} $,  $\mathcal{W}^{-1}=\Pi_{\mathrm{loc}}+U_{ii}^{-1}$ , and  $b_{\pm}=\left(  \mathcal{W}^R(\w={0})-  W_{k=0}^R(\w=0)\right) \aver{\varsigma_i}_{\pm} $, respectively \eqref{eq:n_dmft3},\eqref{eq:n_dmft2},\eqref{eq:n_dmft1} and \eqref{eq:n_dmft4} in this Supplemental. These give new values for $b$ and $\mathcal{W}$, that are transformed back into real times to interate the procedure. 
\end{itemize}

An important point of the update procedure concerns how to define the Green's functions at time $t=0$ such as to interface the NCA impurity solver discussed below and Keldsyh field theory. This will be discussed in Sec. \ref{sec:t=0Greens}. 


\section{NCA impurity solver in the Nambu formalism}
\label{app:nca}

The real-time dynamics of the impurity problem is computed using an impurity solver based on a non-crossing approximation for the dynamical map $\hat{\mathcal{V}}(t)$ that evolves the density matrix $\rho(t) = \hat{\mathcal{V}}(t) \rho(0)$ \cite{schiroScarlatella2019,scarlatellaSchiro2021,scarlatellaSchiro2024a}.  
This approach corresponds to a first-order self-consistent scheme (equivalent to a non-crossing approximation, NCA), obtained by expanding in the system-bath coupling (hybridization expansion). 
The dynamics is obtained by solving the Dyson equation for the superoperator $\hat{\mathcal{V}}(t)$
\begin{equation}
\label{eq:ncaMaps}
\partial_t \hat{\mathcal{V}}(t)= \hat{\mathcal{H}}_{\rm eff} \hat{\mathcal{V}}(t)+\int_0^t d t_1 \hat{\mathcal{D}}\left(t-t_1\right) \hat{\mathcal{V}}\left(t_1\right)
\end{equation}
where $\hat{\mathcal{H}}_{\rm eff} = - i \lsq H, \bullet \rsq + i \sum_{\alpha \beta a b}\varsigma_{\alpha a}^\dagger [\tau_3]_{\alpha a,\beta b} b_{\beta b}$, with the NCA self-energy given by 
\begin{equation}
\label{eq:ncaDissip}
\begin{aligned}
\hat{\mathcal{D}}\left(t\right)=&-\frac{i}{2} \sum_{\alpha \beta a b} \alpha \beta\left[ \mathcal{W} _{b a}^{\beta \alpha}\left(t\right) \hat{\varsigma}_{\beta b}^{\dagger} \hat{\mathcal{V}}\left(t\right) \hat{\varsigma}_{\alpha a}+ \mathcal{W} _{a b}^{\alpha \beta}\left(-t\right) \hat{\varsigma}_{\beta b} \hat{\mathcal{V}}\left(t\right) \hat{\varsigma}_{\alpha a}^{\dagger}\right] = \\ 
&= \frac{1}{2} \sum_{\alpha \beta a b} \alpha \beta (-i) \mathcal{W} _{b a}^{\beta \alpha}\left(t\right) \hat{\varsigma}_{\beta b}^{\dagger} \hat{\mathcal{V}}\left(t\right) \hat{\varsigma}_{\alpha a} + \hc 
\end{aligned}
\end{equation}

Here we introduced the vectors of spin super-operators $\hat{\varsigma} = (\sigma^- \bullet,\sigma^+ \bullet,\bullet\sigma^-,\bullet \sigma^+)^T$ and $\hat{\varsigma}^\dagger = (\sigma^+ \bullet,\sigma^- \bullet,\bullet\sigma^+,\bullet \sigma^-)$, where the bullets indicate whether the corresponding operators apply from the left or from the right, that are analogous to the Nambu-Keldysh vectors of fields in the Keldysh formalism. We refer to their components, i.e. $\hat{\varsigma}_{\alpha a}$, using a Keldysh index $\alpha = +,(-)$ addressing the first(second) half of the vector and a Nambu index $a = 1(2)$ taking the corresponding first (second) component. An analogous notation is used for the other symbols.  
The second equality is obtained by using that the baths correlation functions at time $-t$ are contained in the Nambu components at $t$, and are found using the steady-state relation \eqref{eq:ssCorrFuncs}.

Note that here all quantities are written as functions of a single time assuming a time-independent stationary-state.
Note also that the $(2,2)$ Nambu component gives a contribution to the dissipator equal to the $(1,1)$ component, since $\mathcal{W}_{22}^{\beta \alpha}(t)= \mathcal{W}_{11}^{\alpha \beta}(-t)$. Analogously, the two contributions given by the $(1,2)$ and $(2,1)$ components are equal to each other by \eqref{eq:ssCorrFuncs}. This justifies the overall $1/2$ in \eqref{eq:ncaDissip}.

Correlation functions are computed using the generalized quantum regression theorem as in \cite{scarlatellaSchiro2021}, that in our steady-state case and for $t\geq 0$ reads 
\begin{equation}
\label{eq:ncaCorrFunc}
\begin{aligned}
\chi_{a b}^{\alpha \beta}\left(t \right) =  -i\operatorname{tr}\left[\hat{\zeta}_{\alpha a} \hat{\mathcal{V}}\left(t \right) \hat{\zeta}_{\beta b}^{\dagger} \rho_s \right] + i \tr \lsq \hat{\zeta}_{\alpha a} \hat{\mathcal{V}}\left(t \right) \rho_s \rsq \tr \lsq \hat{\zeta}_{\beta b}^{\dagger} \rho_s \rsq
\end{aligned}
\end{equation}
while negative times are computed by \eqref{eq:ssCorrFuncs}. 

Integrating \eqref{eq:ncaMaps}, \eqref{eq:ncaDissip} needs using an implicit discretization scheme for numerical stability. We use here a second-order Runge-Kutta scheme \cite{scarlatellaSchiro2024a}. Note that this needs to be adapted as to capture correctly a $~\delta(t)$ contribution in the Weiss field $\mathcal{W}$ \cite{scarlatellaSchiro2024a}. 

We remark that in the limit of independent atoms the impurity problem becomes equivalent to a single-site of the original lattice problem, described by a Lindblad master equation. In this limit our NCA approach exactly reduces to a Lindblad master equation \cite{scarlatellaSchiro2024a}, introducing no further approximations.

\section{Definition of Green's functions at t=0}
\label{sec:t=0Greens}

The Green's functions defined on the Keldysh contour don't have a uniquely defined $t\rw 0$ limit: from the definition of time-ordering they satisfy the relations 
\begin{align}
\label{eq:neqGreenProp}
\mathcal{W}^{++} &= \mathcal{W}^{-+} & \mathcal{W}^{--} &= \mathcal{W}^{+-} & &\rm{for\quad} t>0 \\
\mathcal{W}^{++} &= \mathcal{W}^{+-} & \mathcal{W}^{--} &= \mathcal{W}^{-+} & &\rm{for\quad} t<0 
\end{align}
from which one sees that the $t\rw 0^+$ and $t\rw 0^-$ limits of the $++$ and $--$ components are different. 

When discretizing time for numerics though, there's the problem of how to define the Green functions at $t=0$.
This affects the evaluation of the DMFT equations, that consist of a series of convolutions in time or matrix products in frequency,
where the value strictly at $t=0$ would not be important, unless there's $\delta(t)$ contributions (frequency-independent contributions) to the Green functions.
Since we study a Markovian problem, Markovian contributions are always generated: for example a natural initial condition for the effective bath $\mathcal{W}$ is taking $\mathcal{W}(\omega) = W_{r,r'}$ frequency independent, that encodes the local physical Markovian dissipator plus the instantaneous dipole-dipole interaction term. 
In this case, it is necessary to define appropriately the Green's functions at $t=0$, such that the ``causality'' structure of Keldysh Green's functions, i.e. the triangular structure of Green's functions in the basis of classical and quantum fields \cite{kamenevKamenev2023}, is preserved by the DMFT equations. 

In addition, we have to be able to resolve the $t\rightarrow 0^\pm$ limits of the Green functions, which are still a necessary input for the NCA self-energy evaluated at $t=0$ \eqref{eq:ncaDissip}. These cannot be extrapolated from finite times, but must be deduced from the $t=0$ value of the Green's functions.
For this reason we resort to a parametrization that allow to reconstruct the $t \rw 0^\pm$ limits from the $t=0$ value and vice-versa. 

\subsection{t=0 contribution to the NCA dissipator}

The Markovian component of NCA self-energy \eqref{eq:ncaDissip} is
\begin{equation}
\begin{aligned}
\hat{\mathcal{D}}\left(t = 0 \right)=&-\frac{i}{2} \sum_{\alpha \beta a b} \alpha \beta\left[ \mathcal{W} _{b a}^{\beta \alpha}\left(t\rw 0^+\right) \hat{\varsigma}_{\beta b}^{\dagger}  \hat{\varsigma}_{\alpha a}+ \mathcal{W} _{a b}^{\alpha \beta}\left(t\rw 0^-\right) \hat{\varsigma}_{\beta b} \hat{\varsigma}_{\alpha a}^{\dagger}\right]
\end{aligned}
\end{equation}
using $\hat{\mathcal{V}}\left(0\right) =\id $, where the two separate limits of $\mathcal{W}$ from positive and negative times enter. For the Nambu components $\mathcal{W}_{12}$ and  $\mathcal{W}_{21}$ those limits are equal (from their definition), while they're different for $\mathcal{W}_{11}$ and $\mathcal{W}_{22}$. Since $\mathcal{W}_{22}$ and $\mathcal{W}_{11}$ are related, we focus on the latter, whose limits can be parametrized as \eqref{eq:neqGreenProp}
\begin{align}
\mathcal{W}_{11}(t\rw 0^+) &= \bma   E_\kappa - i \frac{\kappa}{2} &  E_f - i \frac{f}{2}  \\ 
 E_\kappa - i \frac{\kappa}{2} &  E_f - i \frac{f}{2}  
\ema 
& \mathcal{W}_{11}(t\rw 0^-) &= \bma   -E_f - i \frac{f}{2} &  -E_f - i \frac{f}{2}  \\ 
 - E_\kappa - i \frac{\kappa}{2} &  -E_k - i \frac{\kappa}{2}  
\ema
\end{align}
where the two are related by \eqref{eq:ssCorrFuncs}.


We will read out $\kappa, f, E_\kappa, E_f$ from the Green's functions at $t=0$, defined in the following, but there only the difference $E_\kappa -E_f$ enters there rather than $E_\kappa$ and $E_f$ separately. 
Therefore, we rather use the alternative parametrization:
\begin{align}
\label{eq:t0LimitsWeiss}
\mathcal{W}_{11}(t\rw 0^+) &= \bma   E_\kappa-E_f - i \frac{\kappa}{2} &  - i \frac{f}{2}  \\ 
 - i \frac{\kappa}{2} &   - i \frac{f}{2}  
\ema 
& \mathcal{W}_{11}(t\rw 0^-) &= \bma   - i \frac{f}{2} &  - i \frac{f}{2}  \\ 
 - i \frac{\kappa}{2} &  -E_k+E_f - i \frac{\kappa}{2}  
\ema
\end{align} 
Using the spin anti-commutation relations, one can check that both parametrizations yields the same $t=0$ value of the $11$ Nambu component of the self-energy, defined by $\hat{\mathcal{D}} = \sum_{ab}\hat{\mathcal{D}}_{ab}$, that reads
\begin{equation}
\label{eq:t=0Diss}
\es{
\hat{\mathcal{D}}_{11}(t=0) \rho &= \oh \lsq  \lp \frac{\kappa}{2} + i E_\kappa \rp \lp d \rho d^\da - d^\da d \rho \rp + \lp \frac{f}{2} + i E_f \rp \lp d^\da \rho d - d d^\da \rho \rp + \hc \rsq = \\
&= \oh \lbr  - i E_\kappa \lsq d^\da d, \rho \rsq -i E_f \lsq d d^\da , \rho \rsq + \lsq \frac{\kappa}{2} \lp d \rho d^\da - d^\da d \rho +\hc \rp  + \frac{f}{2} \lp d^\da \rho d - d d^\da \rho +\hc \rp \rsq \rbr
}
\end{equation}
where we called $d =\sigma^-$ to avoid confusion with Keldysh indices.
Then $\mathcal{W}_{22}(t\rightarrow 0^+)=  \mathcal{W}^T_{11}(t\rightarrow 0^-)$ is obtained from the relation $\mathcal{W}(t)_{22}^{\alpha \beta} = \mathcal{W}(-t)_{11}^{\beta \alpha}$. 

\subsection{$t=0$ definition of Keldysh Green functions}

The proper definition of a Keldysh action in the presence of a Markovian environment, restricting to the 11 Nambu component, is 
\cite{siebererDiehl2016,siebererDiehl2014,kamenevKamenev2023}.
\begin{equation}
S_{11}= - \oh \int_{-\infty}^\infty d t^\prime
\bma \vdots \ema^\da 
\tau_3
{\mathcal{W}}_{11}(t=0)
\tau_3
\bma 
d_+ \\ d_- 
\ema
\end{equation}
with the Green function of the Markovian environment parametrized as
\begin{equation} 
\label{eq:t=0ActWeiss}
{\mathcal{W}}_{11}(t=0) = \bma
E_k - i \frac{\kappa}{2} {- E_f - i \frac{f}{2}} & -i f \\
-i \kappa & {-E_k -i\frac{\kappa}{2}} +E_f - i \frac{f}{2}
\ema
\end{equation}
A direct mapping exists between a Markovian master equation and an action of this form \cite{siebererDiehl2016,siebererDiehl2014,kamenevKamenev2023}, that identifies the coefficients in Eq. \eqref{eq:t=0ActWeiss} with those in Eq. \eqref{eq:t0LimitsWeiss}.
Note that \eqref{eq:t=0ActWeiss} simply corresponds to $\mathcal{W}_{11}(t=0) = \mathcal{W}_{11}(t \rw 0^+) + \mathcal{W}_{11}(t\rw 0^-)$, with the definition of \eqref{eq:t0LimitsWeiss}.
Also note that \eqref{eq:t=0ActWeiss} enforces the ``causal'' structure of the Keldsyh action, corresponding to the triangular structure when a rotation to the classical-quantum basis is performed:
\begin{equation} 
{\mathcal{W}}_{11}(t=0) \rw \bma
- i ( {\kappa} + f ) & -i \frac{\kappa - f}{2} + E_k -E_f \\ 
i \frac{\kappa -f }{2} + E_k -E_f & 0 
\ema
\end{equation}

Finally, the $22$ Nambu component of the action is simply related to the $11$ component by ${\mathcal{W}}_{22}(t=0)={\mathcal{W}}_{11}(t=0)^T$. Instead the $12$ and $21$ terms in the action are ${\mathcal{W}}_{12(21)}(t=0)= 2 {\mathcal{W}}_{12(21)}(t \rw 0^\pm) $ since the 2 limits are equal and they add up in the mapping.

\subsection{t=0 component in DMFT iterations}

In practice, in the numerical iterative scheme to solve the DMFT \ref{sec:numImp} equations, the following steps must be added for the $t=0$ component: 
\begin{align}
\rightarrow {\mathcal{W}}(0) \quad {\longrightarrow} \quad \mathcal{W}(0^+),\mathcal{W}(0^-) \underset{\rm NCA\,impurity\,solver }{\longrightarrow} \chi(0^+)\chi(0^-) \longrightarrow \quad {\chi}(0) \quad \underset{\textrm{DMFT\;\;equations}}{\rw}  
\end{align}
Namely, before passing the effective environment ${\mathcal{W}}$ to the NCA impurity solver, the $t\rw 0^{\pm}$
limits are resolved, extracting the parameters $E_\kappa -E_f$, $\kappa$ and $f$ from ${\mathcal{W}}(t=0)$ and using them to determine $\mathcal{W}(0^-)$ and $\mathcal{W}(0^+)$ with \eqref{eq:t0LimitsWeiss}. 
Then, the atom correlation function $\chi(0^\pm)$ is computed by the NCA solver \eqref{eq:ncaCorrFunc} and $\chi(0)$ is defined by \eqref{eq:t=0ActWeiss}, before being inputted in the DMFT equations.

\putbib
\end{bibunit}

\end{document}